\documentclass{moriond}

\def\be{\begin{equation}}
\def\ee{\end{equation}}
\def\bea{\begin{eqnarray}}
\def\eea{\end{eqnarray}}
\def\$CP$bar{\hbox{{\rm $CP$}\hskip-1.80em{/}}}

\RequirePackage{xspace}
\def\Bbar    {\kern 0.18em\overline{\kern -0.18em B}{}\xspace}

\def\Kbar    {\kern 0.18em\overline{\kern -0.18em K}{}\xspace}
\def\Kb      {\ensuremath{\Kbar}\xspace}
\newcommand{\KorKbar}{\raisebox{7.7pt}{$\scriptscriptstyle(\hspace*{9.9pt})$}
  \hspace*{-13.4
pt}\Kbar{}^{\,\, 0}\,}

\newcommand{\Photo}{\includegraphics[height=35mm]{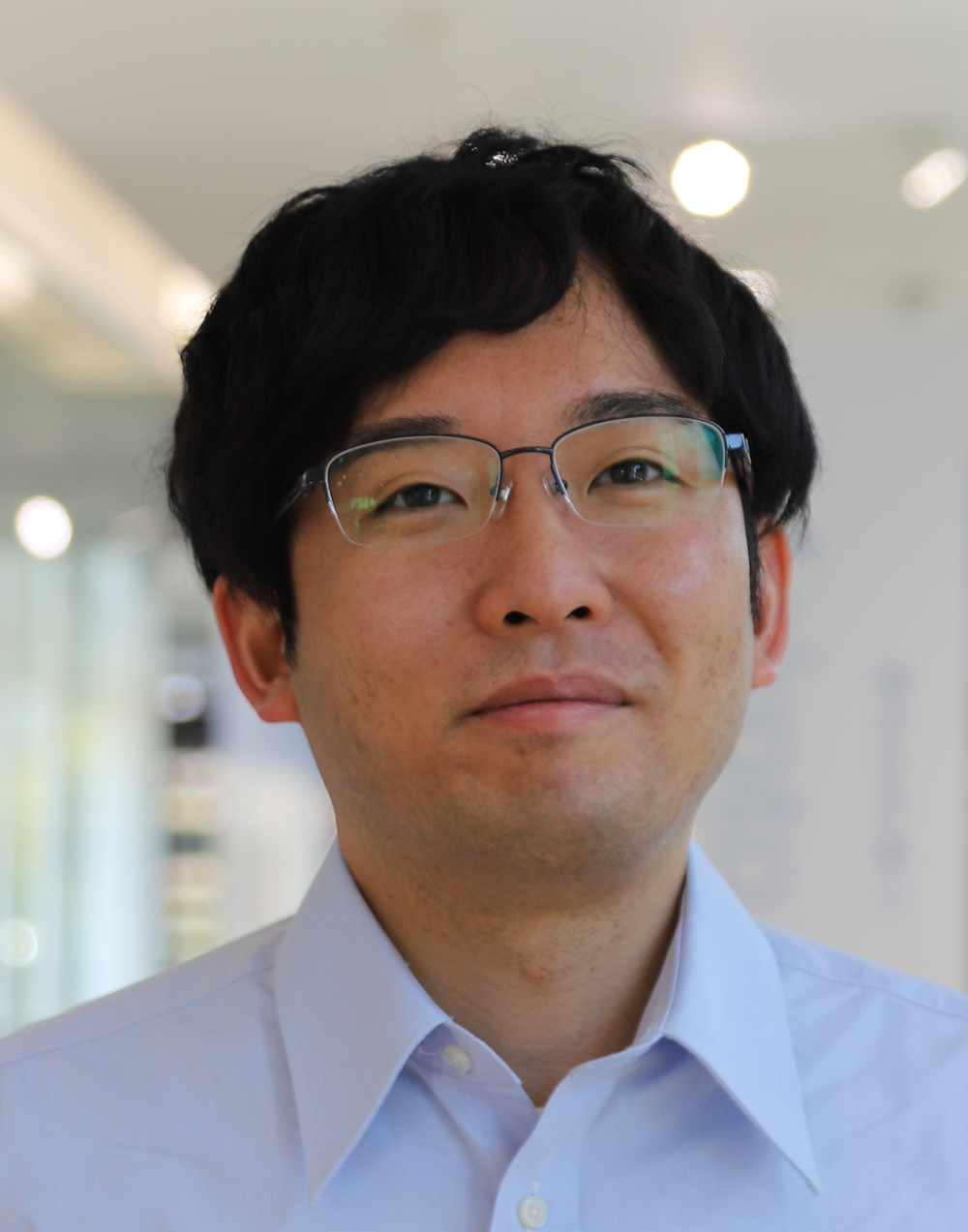}}

\hypersetup{
  colorlinks=true,
  linkcolor=blue,
  citecolor=blue,
  urlcolor=blue
}

\usepackage{amsmath}
\usepackage{cite}
\usepackage{comment}

\begin{document}
\hfill CHIBA-EP-279\\
\vspace*{4cm}
\
\title{New Directions in Kaon Physics:\\[0.15em]
Interference in \boldmath{$K^0\to\mu^+\mu^-$} as a New Golden Mode
\footnote{Contribution to the Proceedings of the 60th Rencontres de Moriond
``Electroweak Interactions \& Unified Theories'', La Thuile, Italy, 15-22 March, 2026.}}

\author{Teppei Kitahara}

\address{Department of Physics, Graduate School of Science,
Chiba University, Chiba 263-8522, Japan\\
Kobayashi-Maskawa Institute for the Origin of Particles and the
  Universe, \\
  Nagoya University,
  Furo-cho Chikusa-ku, Nagoya 464-8602, Japan\\
        E-mail: \href{mailto:kitahara@chiba-u.jp}{kitahara@chiba-u.jp}}

\maketitle\abstracts{The rare decays $K^0_L \to\mu^+\mu^-$ and $K^0_S \to\mu^+\mu^-$ have long been regarded 
as difficult channels to extract short-distance physics because they are dominated by long-distance contributions via two-photon exchanges.
A qualitatively new feature arises once the interference between $K_L^0$ and $K_S^0$ is taken into account.
The interference term is sensitive to genuine direct $CP$ violation in  $s \to d \mu^+ \mu^-$ processes and turns this channel into a clean probe of the $CP$-violating short-distance physics.
In this contribution,
I summarize the basic mechanism of the $K_L^0$--$K_S^0$ interference,
 the flavor-tagging strategy at LHCb-like setup,
  and the projected sensitivities for a kaon-unitarity-triangle parameter combination $|A^2\lambda^5\bar\eta|$ and also for the sign ambiguity of the $K_L^0 \to \gamma \gamma$
  three-point amplitude.  
  As a result, it is expected that 
   the CKM parameter $|A^2\lambda^5\bar\eta|$ could be constrained by LHCb at the level of about 35\% of its Standard Model (SM) value, and the discrete ambiguity in $\mathcal{B}(K_L^0 \to \mu^+\mu^-)_{\rm SM}$ could be resolved at more than $3\sigma$ by the end of the high luminosity LHC.
 }

\section{Introduction}

The CKM paradigm is an extremely successful framework, 
providing a unified scheme to describe a vast number of experimental observables in a consistent way. 
Looking ahead, many new endeavors are underway, aiming to improve the precision with which we measure the CKM parameters, 
and in parallel searching for novel tests in uncharted territories. 
One such territory is kaon physics, where rare kaon decays remain among the sharpest laboratories for testing flavor dynamics, 
$CP$ violation, 
and possible new physics beyond the Standard Model (SM). 
The recent observation of $K^+\to\pi^+\nu\bar\nu$ by NA62, 
together with current theory and experimental knowledge of $|\varepsilon_K|$,
has begun to form a CKM determination independent of $B$-physics inputs~\cite{Dery:2025pcx}. 
In this broad program, 
the ultra-rare decay $K^0\to\mu^+\mu^-$ occupies a rather unusual position. 
It is experimentally attractive because of its clean charged final state, 
but theoretically challenging because its decay rates are largely dominated by long-distance physics.

The standard obstacle is well known. 
The decay $K_L^0\to\mu^+\mu^-$ receives a large $CP$-conserving contribution from the two-photon intermediate state $K_L^0\to\gamma^\ast \gamma^\ast \to\mu^+\mu^-$. 
Since this long-distance amplitude is numerically significant, 
the total rate alone is not an optimal observable for short-distance $CP$-violating physics. 
In recent years, however, $K^0\to\mu^+\mu^-$ has profited from substantial theory progress. The key observation of Refs.~\cite{DAmbrosio:2017klp,Dery:2021mct} is that one can bypass the challenge of long-distance physics estimations altogether by focusing on interference effects between $K_L^0$ and $K_S^0$ decays to a $\mu^+\mu^-$ pair. 
These interference effects are, to a very good approximation, 
dictated solely by short-distance contributions and can therefore be used to cleanly extract CKM parameters.

This idea has now reached a stage where an experimental assessment is possible. 
In Ref.~\cite{DAmbrosio:2025mxa}, a fast simulation of a LHCb-like setup was carried out by combining the interference formalism with realistic flavor tagging, 
background rejection, and decay-time information. 
The analysis uses both time-integrated and time-dependent information to extract the parameters of interest. 
The conclusion is that $K^0\to\mu^+\mu^-$ may become a genuine precision observable in the LHCb Upgrade era, with implications both for kaon phenomenology and for independent tests of the CKM paradigm from kaon physics.

\section{Why the $K_L^0$--$K_S^0$ interference is special}

The relevant $\mu^+\mu^-$ final state contains both $CP$-odd and $CP$-even components. 
For unpolarized muons, 
the $s$-wave spin-singlet dimuon configuration,
$(\ell,S)=(0,0)$, is $CP$-odd, 
whereas the $p$-wave spin-triplet configuration,
$(\ell,S)=(1,1)$, is $CP$-even, 
since the total angular momentum of the kaon is zero.
For an initial $K^0$ or $\Kb^0$ beam, 
the time-dependent rate is written~as
\begin{eqnarray}\label{eq:timeDep}
\frac{1}{{\cal N}}\frac{d\Gamma(\KorKbar \to\mu^+\mu^-)}{dt} \,  = C_L \,e^{-\Gamma_L t} + C_S \,e^{-\Gamma_S t} \pm  2\,C_\text{Int.}\cos(\Delta M_K t-\varphi_0) e^{-\Gamma t}\, ,
\end{eqnarray}
where $\Gamma=(\Gamma_L+\Gamma_S)/2$. 
Here $C_L$ and $C_S$ describe the pure $K_L$ and $K_S$ contributions, while $C_{\rm Int.}$ and $\varphi_0$ encode the magnitude and phase of the $K_L^0$--$K_S^0$ interference. 
The key point is that $C_{\rm Int.}$ determines the short-distance $\ell=0$ amplitude through $|A(K_S)_{\ell=0}|^2=C_{\rm Int.}^2/C_L$ \cite{Dery:2021mct}.

The interference term provides a clean handle on direct $CP$ violation in $K^0\to\mu^+\mu^-$. 
The long-distance contribution to $K_L^0\to\mu^+\mu^-$ is dominated by the two-photon intermediate state with single virtual $\pi^0$ exchange, whose the strong-phase ($CP$-even phase) is large, 
whereas the short-distance amplitude carries the weak phase ($CP$-odd phase), proportional to $V_{td}V_{ts}^*$ within the SM. 
Therefore, the interference term is directly sensitive to the product of the relative phases of the $CP$-even and $CP$-odd phases.
Equivalently, the time-integrated $CP$ asymmetry, $\widetilde A_{\rm CP}$, is proportional to $C_{\rm Int.}\,I_{\rm Int.}$, normalized by the sum of the $K_L^0$ and $K_S^0$ contributions. 
Since the efficiency-convoluted interference integral can be determined once the time acceptance is known, 
the measurement of the asymmetry gives access to the short-distance amplitude.
Note that the detailed definitions of $C_{\rm Int.}$ and $I_{\rm Int.}$ are given in Ref.~\cite{DAmbrosio:2025mxa}.

A further important feature is the discrete ambiguity associated with the long-distance amplitude $A(K_L^0\to\gamma\gamma)$. 
The phase $\varphi_0$ is known only up to a four-fold ambiguity,
 which is related to the unknown overall sign of this amplitude. 
In the relevant time range, the sign of the time-integrated $CP$ asymmetry, $\widetilde A_{\rm CP}$, is controlled by the sign of $\cos\varphi_0$ \cite{Dery:2022yqc}. 
Thus measuring the sign of the asymmetry will resolve the ambiguity and select one of the two SM predictions \cite{Hoferichter:2023wiy},
\begin{align}\label{eq:prediction-KLmumu}
\mathcal{B}\left(K^0_L \rightarrow \mu^{+} \mu^{-}\right)_{\rm{SM}}=\left\{\begin{array}{l}
7.44^{+0.41}_{-0.34}  \times 10^{-9}\qquad \text{for}~ \widetilde A_{\rm CP}>0\,, \\
6.83^{+0.24}_{-0.17}  \times 10^{-9}\qquad \text{for} ~\widetilde A_{\rm CP}<0\,.
\end{array}\right.
\end{align}
In this way, the interference pattern not only probes short-distance $CP$ violation, but also removes an important long-distance ambiguity in the SM prediction of $\mathcal{B}(K_L^0\to\mu^+\mu^-)$.

\section{Flavor tagging and background suppression at LHCb}

\begin{figure}[t]
  \centering
\quad  \includegraphics[width=0.8\linewidth]{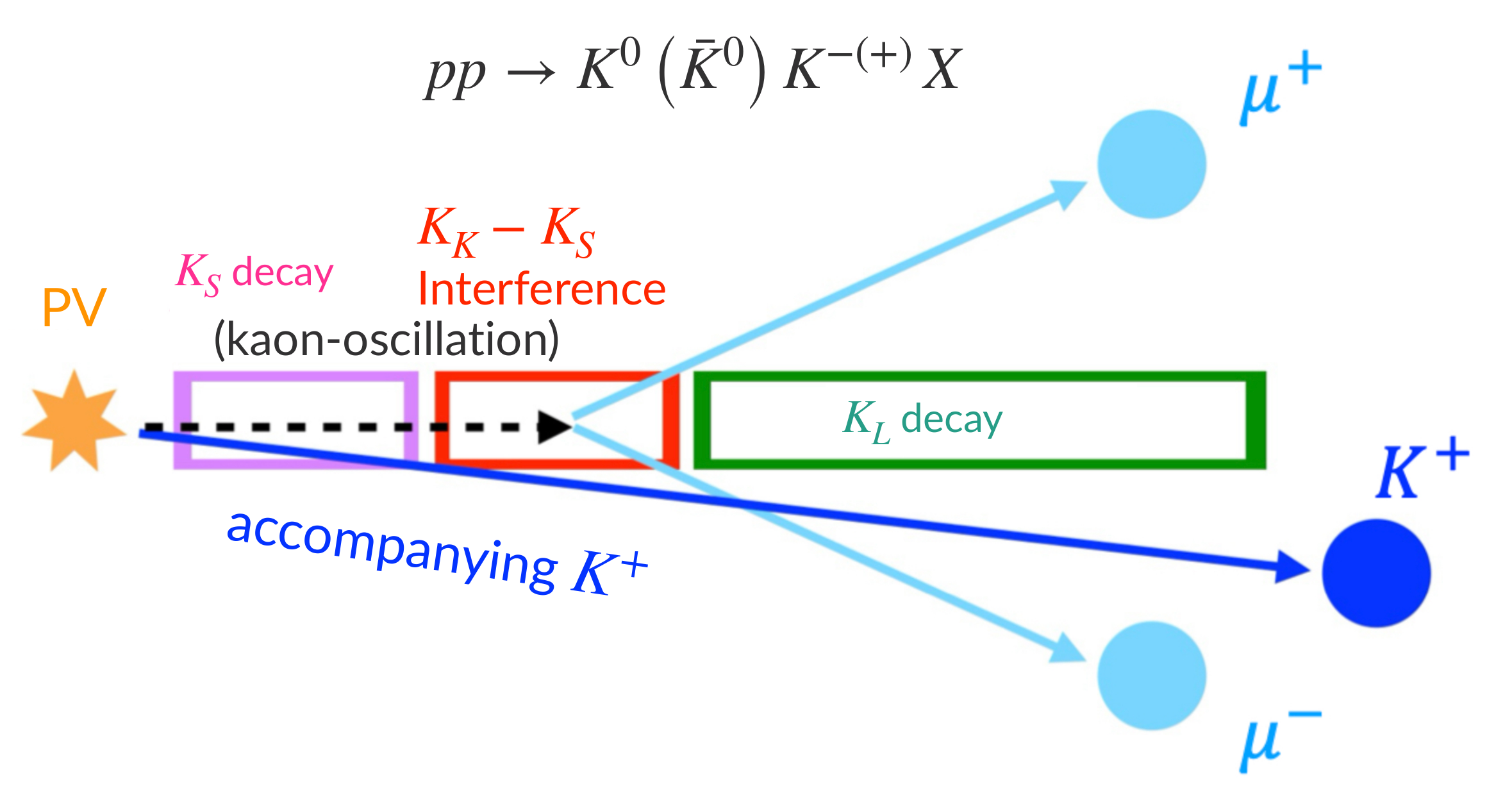}
\vspace{-0.2cm}
  \caption{Schematic picture of the tagged $K^0(\Kb^0) (t) \to \mu^+\mu^-$ analysis considered in the fast simulation. 
  Neutral kaons at the primary vertex produced in 
 $pp \to K^0K^{-}X$ and $pp \to \Kb^0 K^{+}X$ are flavor tagged by the QED charge of the accompanying charged kaon, and they subsequently evolve into a $K_L^0$--$K_S^0$ admixture. 
  The reconstructed decay position provides time-dependent information sensitive to  $CP$ violation.}
  \label{fig:tagging}
\end{figure}

The experimental challenge is to determine whether the neutral kaon produced at the primary vertex was initially $K^0$ or $\Kb^0$. 
The strategy studied in Ref.~\cite{DAmbrosio:2025mxa} uses the associated production process 
\begin{equation}
pp\to K^0\,K^{-}X \quad \textrm{and} \quad  pp\to \Kb^0\,K^{+}X\,,
\label{eq:tagproc}
\end{equation}
where the charge of the accompanying charged kaon provides the flavor tag \cite{DAmbrosio:2017klp}. 
The schematic picture is shown in Fig.~\ref{fig:tagging}.
This is closely analogous to same-side-kaon tagging for $B_s$ mesons \cite{LHCb:2012zja}, but the kaon environment is less busy (significantly smaller $K^\pm$ multiplicity)
 and therefore experimentally more favorable. 
 In the fast simulation in Ref.~\cite{DAmbrosio:2025mxa}, 
 one obtains an efficiency of the tagging algorithm of about $62\%$ and a dilution factor of about $60\%$ (defined by $D = 1 - 2 \omega$ with the mistag rate $\omega$),
  corresponding to a tagging power
\begin{equation}
T_P =\varepsilon_T D^2 \approx 22\%\,.
\label{eq:TP}
\end{equation}
This is much larger than the tagging power typical of same-side-kaon tagging in the $B_s$ system.

The dominant background is currently expected to come from
\begin{equation}
K_S^0\to \pi^+(\to\mu^+\nu)\,\pi^-(\to\mu^-\bar\nu)\,,
\label{eq:bkg}
\end{equation}
where at least one of the pions decayed in flight to a muon and a neutrino \cite{LHCb:2020ycd}.
It is shown in Ref.~\cite{DAmbrosio:2025mxa} that
 a very useful selection is the impact parameter of 
 the reconstructed invisible $K^0$ line with respect to the primary vertex.
  For a genuine signal, this impact parameter is consistent with zero up to the detector resolution, 
  while the background can take wider ranges.
 The analysis of Ref.~\cite{DAmbrosio:2025mxa} found that an impact-parameter cut improves the signal sensitivity by roughly a factor of $9$.
In the most optimistic scenarios, 
 the background can be reduced by one or two orders of magnitude by sacrificing about
half of the efficiency.

To first approximation, the decay-time acceptance for $K^0\to\mu^+\mu^-$ can be described by an allowed decay-time range, driven by selection cuts, together with an exponential acceptance factor~\cite{AlvesJunior:2018ldo},
$
F^{\rm eff}(t)\simeq
\Theta(t-t_{\min})\Theta(t_{\max}-t)\,e^{-\beta_{\rm acc.} t}.
$
The parameters entering this effective acceptance are strongly dependent on the trigger, selection, and tracking configuration.
In particular, the use of downstream $K^0\to\mu^+\mu^-$ decays in the Upgrade~II scenario \cite{LHCb:2021glh}, enabled by the improved angular and invariant-mass resolution of the Upstream Pixel detector, can enlarge the accessible decay-time range and increase the signal yield and further increase the signal yield by up to a factor of $3$ \cite{DAmbrosio:2025mxa}.
Thus, the same detector features that improve the decay-time acceptance also will help control decays-in-flight backgrounds, making the measurement especially well matched to the LHCb geometry and to boosted neutral kaons.

\section{Decay-time analysis and projected sensitivities}

\begin{figure}[t]
  \centering
  \includegraphics[width=0.8\linewidth]{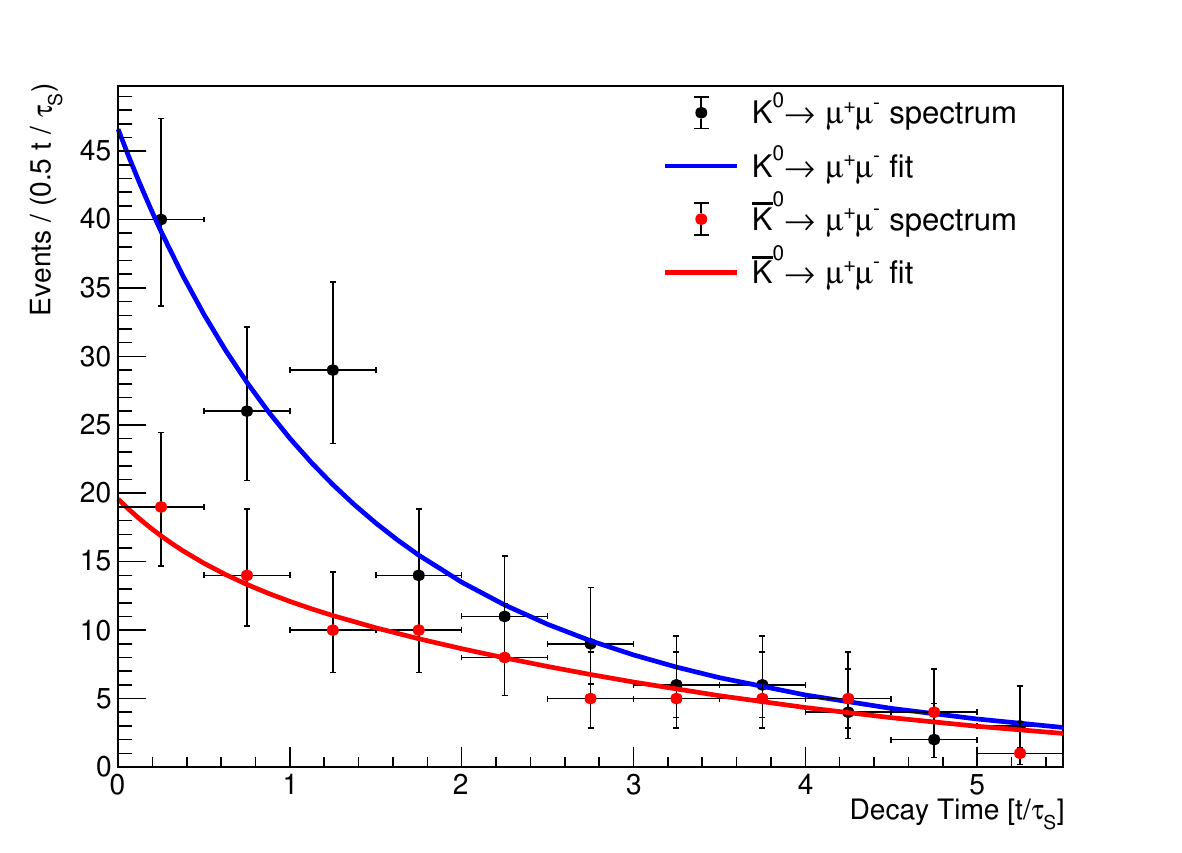}
  \caption{Decay-time distributions for $K^0\rightarrow\mu^+\mu^-$ (blue) and $\Kb^0\rightarrow\mu^+\mu^-$ decays (red), using the decay-time acceptance for the Upgrade II scenario. The functions after the unbinned simultaneous fit are overlaid. The spectra are shown for $Y_{eff} \sim 300$ events.}
    \label{fig:exp_decay-time}
\end{figure}

Once the initial kaon flavor is tagged, the time dependence of the dimuon signal carries the information on the $K_L^0$--$K_S^0$ interference. 
 The simulated decay time distributions  for the Upgrade II scenario are shown in Fig.~\ref{fig:exp_decay-time}.
 It is clearly shown that the $CP$ violation between $K^0 \to \mu^+ \mu^-$ (accompanying $K^-$) 
 and
 $\Kb^0 \to \mu^+ \mu^-$ (accompanying $K^+$) 
 can be
measured for small decay time region. 

\begin{figure}[t]
  \centering
  \includegraphics[width=0.8\linewidth]{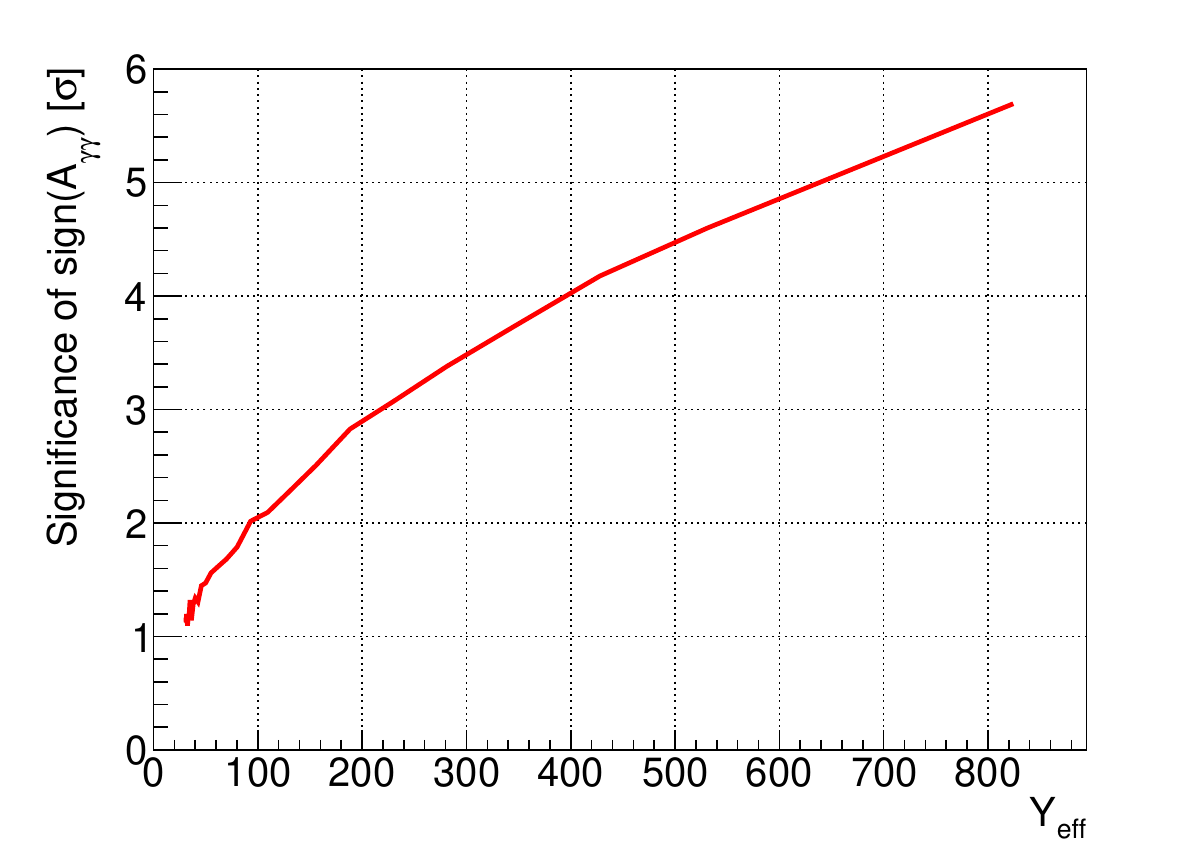}
  \caption{Experimental sensitivity to  the significance of the sign of $A_{\gamma\gamma}$ assuming the SM, as a function of the effective yield $Y_{eff}$. 
  It can be seen that with $Y_{eff} \sim 300$ events, LHCb can resolve ${\rm sign}[A_{\gamma\gamma}]$ at more than three standard deviations.}
  \label{fig:sign}
\end{figure}

Two interesting projected sensitivities have been investigated in the  $CP$-violating  observables. 
The first is the unknown overall sign of the amplitude $A(K_L^0\to\gamma\gamma)$. 
A simultaneous fit to the tagged decay-time spectra can distinguish the discrete ambiguity present in the SM prediction in Eq.~\eqref{eq:prediction-KLmumu}.
The expected sensitivity is summarized in Fig.~\ref{fig:sign}
as a function of the effective yield 
$Y_{eff} = T_P \, S_{eff}$, where 
$S_{eff} = S^2/(S +B) \times (\textrm{improvement factors}) $ denotes 
the purity-corrected signal yield.
In the luminosity range relevant for LHCb Run~5/6, 
the eventual effective yield is expected to exceed $Y_{\rm eff} \sim 300$ events.
At this level, the significance reaches the threshold required to resolve the sign ambiguity, thereby removing the corresponding discrete theoretical ambiguity in
$\mathcal{B}(K_L^0\to\mu^+\mu^-)$.

\begin{figure}[t]
  \centering
  \includegraphics[width=0.8\linewidth]{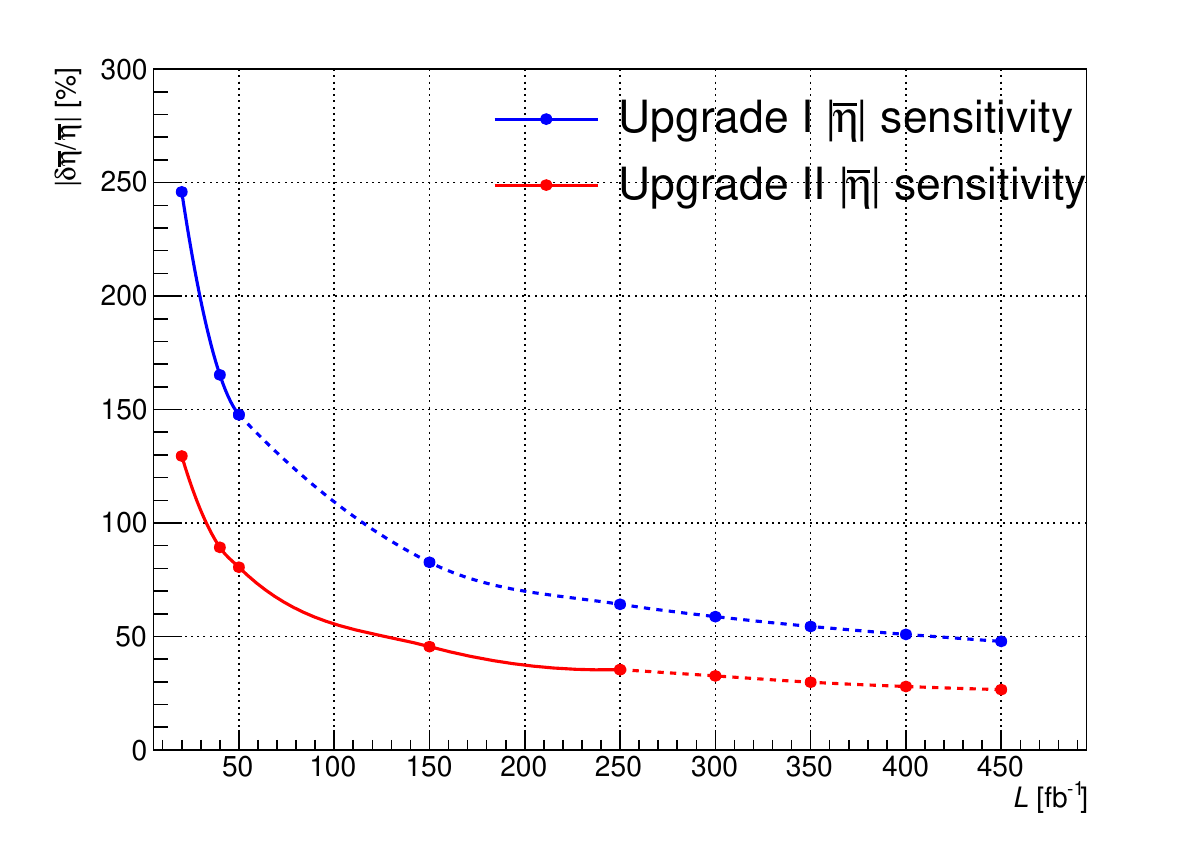}
  \caption{Sensitivity to  $\vert\bar{\eta}\vert$ as a function of the integrated luminosity for Upgrade~I (Upgrade~II) scenarios with $T_P=22\%$, $S_{eff}^{DD}/S_{eff}^{LL} = 0\,(1.9)$ and a $S^2/(S+B)$ improvement factor of $\times9$.
    Solid lines correspond to currently considered expected luminosities, while the dashed lines demonstrate the behavior outside of the current plans and serve to compare between the performances in the two considered scenarios. }
  \label{fig:eta}
\end{figure}

The second target is the $CP$-violating parameter $|A^2\lambda^5\bar\eta|$,  
which directly connects the interference measurement to the kaon unitarity triangle.
As summarized in Fig.~\ref{fig:eta} as a function of the integrated luminosity, 
the LHCb Upgrade~I and Upgrade~II are expected to reach accuracies of about $150\%$ and $35\%$, respectively.
The improvement of the acceptance enabled by the planned Upstream Pixel detector
is essential to reach this best sensitivity.
Also, significant improvements on background rejection are needed, but seem possible.

The phenomenological meaning is particularly transparent in the kaon unitarity-triangle.
A useful point emphasized recently is that one can consider the triangle not in the usual $(\bar\rho,\,\bar\eta)$ plane,
 but in the rescaled plane $(A^2(1-\hat\rho),\,A^2\hat\eta)$ with 
 $A^2 \lambda^4(1-\hat{\rho}+i \hat{\eta}) = -V_{t d} V_{t s}^*/V_{c d} V_{c s}^*$, 
where the uncertainty from $|V_{cb}|$ 
is effectively removed from the unitarity triangle~\cite{Dery:2025pcx}.
In this language, the interference observable directly probes the vertical coordinate through $A^2\lambda^5\bar\eta$ ($\hat{\eta}$ coincides to $\mathcal{O}(\lambda^2)$ with $\bar{\eta}$). 
This makes $K^0\to\mu^+\mu^-$ interference a valuable complement to rare-neutrino modes such as $K_L^0\to\pi^0\nu\bar\nu$.
Although both are sensitive to the same short-distance CKM structure, 
the dimuon mode provides qualitatively different information through a tagged time-dependent asymmetry and, at the same time, can probe the sign of the two-photon amplitude.

\section{Conclusions}

The outstanding particle yields of the current era, together with upgraded detector capabilities, open the door to new directions in rare kaon decays. 
In this context, $K^0\to\mu^+\mu^-$ is emerging as a new golden mode, 
where the $K_L^0$--$K_S^0$ interference turns a long-distance-dominated process into a powerful probe of direct $CP$-violating short-distance physics.
The key ingredient is the interference between the enhanced long-distance amplitude $K_L^0\to\gamma^\ast\gamma^\ast\to\mu^+\mu^-$ and the short-distance contributions with the weak phase, which appears as a measurable effect in tagged, time-dependent distributions, and also in time-integrated $CP$ asymmetry.

The LHCb-like study in Ref.~\cite{DAmbrosio:2025mxa} shows that
 such a measurement is experimentally challenging 
  but realistic under benchmark performance assumptions. 
In the most optimistic Upgrade~II scenario, 
the CKM parameter $|A^2\lambda^5\bar\eta|$ 
could be determined at the level of roughly $35\%$, 
while the sign ambiguity in $A(K_L^0\to\gamma\gamma)$ can be resolved with a significance above $3\sigma$. 
This would provide a valuable complement to the determination of the same CKM structure through $K_L^0\to\pi^0\nu\bar\nu$ at the KOTO and proposed KOTO~II experiments, while also eliminating a discrete ambiguity in the SM prediction for $\mathcal{B}(K_L^0\to\mu^+\mu^-)$. 
Thus, $K^0\to\mu^+\mu^-$ interference offers a novel opportunity to test the kaon unitarity triangle and deserves dedicated experimental studies within LHCb.

\section*{Acknowledgments}

I would like to thank 
Giancarlo D'Ambrosio, 
Vital Dery, 
Yuval Grossman, 
Radoslav Marchevski, 
Diego Martínez Santos, and 
Stefan Schacht  for fruitful collaborations on the presented work. 
I would also like to warmly thank the organizers of Moriond EW 2026 for inviting me to present these results at such an excellent place.
This work is supported by the JSPS Grant-in-Aid for Scientific Research Grant No.\,22K21347, 24K22872 and 25K07276. 

\section*{References}


\end{document}